\documentclass[10pt]{article}
\usepackage{float}
\usepackage[caption = false]{subfig}
\usepackage[final]{graphicx}

\def\Title#1{\begin{center} {\Large #1 } \end{center}}
\def\Author#1{\begin{center}{ \sc #1} \end{center}}
\def\Address#1{\begin{center}{ \it #1} \end{center}}

\newcommand\pubblock{\rightline{\begin{tabular}{l} Proceedings of the Fifth Annual LHCP\\ \pubnumber\\
         \pubdate  \end{tabular}}}

\newenvironment{Abstract}{\begin{quotation} \begin{center} 
             \large ABSTRACT \end{center}\bigskip 
      \begin{center}\begin{large}}{\end{large}\end{center} \end{quotation}}

\newenvironment{Presented}{\begin{quotation} \begin{center} 
             PRESENTED AT\end{center}\bigskip 
      \begin{center}\begin{large}}{\end{large}\end{center} \end{quotation}}

\def\beq{\begin{equation}}
\def\eeq#1{\label{#1}\end{equation}}
\def\eeqn{\end{equation}}

\def\beqa{\begin{eqnarray}}
\def\eeqa#1{\label{#1}\end{eqnarray}}
\def\eeqan{\end{eqnarray}}

\let\bar=\overbar

\def\Dslash{\not{\hbox{\kern-4pt $D$}}}
\def\dslash{\not{\hbox{\kern-2pt $\del$}}}

\def\msb{{\bar{\ssstyle M \kern -1pt S}}}

\usepackage{color}

\newcommand\pubnumber{}

\newcommand\pubdate{\today}

\def\affiliation{
On behalf of the CMS and ATLAS Collaborations, \\
Institute of High Energy Physics\\
19B Yuquan Lu, Shijingshan District, Beijing, 100049, China}

\begin{document}

\large
\begin{titlepage}
\pubblock

\vfill
\Title{  Searches for new physics in lepton plus jet final states\\ in ATLAS and CMS  }
\vfill

\Author{ Francesco Romeo  }
\Address{\affiliation}
\vfill
\begin{Abstract}
The most recent results on searches in lepton plus jet final states motivated by different models beyond the standard model are presented,
using $pp$ collision data collected by the ATLAS and CMS detectors during Run I and Run II at the CERN LHC.
Leptoquarks and heavy Majorana neutrinos that arise in the left-right model
are looked for in the final states with two leptons and two jets ($\ell\ell + jj, \ell = e,\mu,\tau$).
Heavy Majorana neutrinos are further investigated by relying on a composite-fermion scenario, considering two leptons and one large-radius jet ($\ell\ell + J, \ell = e,\mu$),
and in the context of a model with Type-1 seesaw mechanism, requiring two same-sign leptons plus dijet ($\ell\ell' + jj, \ell,\ell' = e,\mu$).
Finally, models of microscopic black holes with two to six extra dimensions are tested by analysing 
the channels with at least one lepton and two additional jets ($\ell + jj, \ell = e,\mu$).
In all the searches, the observed data are in good agreement with the standard model prediction
and 95\% confidence level upper limits are set on the parameters of different models.
\end{Abstract}
\vfill

\begin{Presented}
The Fifth Annual Conference\\
 on Large Hadron Collider Physics \\
Shanghai Jiao Tong University, Shanghai, China\\ 
May 15-20, 2017
\end{Presented}
\vfill
\end{titlepage}
\def\thefootnote{\fnsymbol{footnote}}
\setcounter{footnote}{0}
%

\normalsize 

\vspace{-0.5cm}
\section{Introduction}
\vspace{-0.15cm}
In this proceeding we summarize the most recent results on searches for new physics with lepton plus jet final states by the ATLAS \cite{ATLAS} and CMS \cite{CMS} collaborations.
This signature is expected in many extensions of the standard model (SM), of which the scenarios dealing with leptoquarks, heavy neutrinos, and microscopic black holes 
are of particular interest and will be considered below.

\vspace{-0.25cm}
\subsection{Leptoquarks}
\vspace{-0.15cm}
Leptoquarks (LQ) are SU(3) color-triplet bosons that carry both lepton and baryon numbers.
The exact properties (spin, weak isospin, electric charge, chirality of the fermion couplings, and fermion number) depend
on the structure of each specific model by which they are foreseen. For this reason, direct searches for LQs 
rely on the Buchm\"{u}ller-R\"{u}ckl-Wyler (BRW) model \cite{BRW} that includes a general effective lagrangian 
and provides symmetry between leptons and quarks of the SM. 
It foresees that LQs of the n$th$ generation would decay into leptons and quarks of the same generation. 
Pair-produced scalar leptoquarks decaying to two leptons and two quarks 
are considered in the analyses described here, for which the cross-section depends only on the LQ mass.
It is worth highlighting that models with LQ of TeV scale masses have recently gained new interest because of
the possibility to interpret the recent anomalies observed in B meson decays by the LHCb experiment \cite{LQLHCb}.

\vspace{-0.25cm}
\subsection{Heavy neutrinos}
\vspace{-0.15cm}
Heavy neutrinos are foreseen in many theories beyond the SM, among which we consider the left-right (LR) model \cite{lr3}, the seesaw mechanism \cite{SeeSaw}, and the composite scenario \cite{HCMNpheno}.
We analyse the process in which the heavy neutrino ($N_{\ell}, \ell = e, \mu, \tau$) is produced in association with a lepton and subsequently decays into a lepton and two quarks. 

\vspace{-0.25cm}
\subsubsection{Left-right model}
\vspace{-0.15cm}
The LR symmetric extension to the SM explains parity violation in the SM as the consequence of spontaneous symmetry breaking at a multi-TeV mass scale. 
It introduces an additional right-handed $SU(2)_{R}$ symmery group to the SM and includes heavy charged ($W_{R}^{\pm}$) and neutral ($Z_{R}$) gauge bosons
and heavy right-handed Majorana neutrinos.
The production cross-section and decay branching-ratio is ruled by the mass $m_{W_{R}}$ and the ratio between $m_{W_{R}}$ and $m_{N_{\ell}}$.

\vspace{-0.25cm}
\subsubsection{Seesaw mechanism}
\vspace{-0.15cm}
The seesaw mechanism can be realized in different beyond SM frameworks (including the previously mentioned LR extension) to explain
the smallness of the neutrino masses, $m_{\nu}$. In the simplest scheme,  $m_{\nu} \approx y_{\nu}^{2} v^{2} / m_{N_{\ell}}$,
where $y_{\nu}^{2}$ is a Yukawa coupling, $v$ is the Higgs vacuum expectation value in the SM, and $m_{N_{\ell}}$ is the mass of a new heavy Majorana neutrino state. 
Different types of seesaw mechanisms exist depending on how they are generated. The so-called ``Type-1" seesaw mechanism, considered here, is implemented through a fermion singlet.
The production cross-section depends on $m_{N_{\ell}}$ and $V_{\ell N}$, where $V_{\ell N}$ describes the mixing between $N_{\ell}$ and the SM neutrino of flavour $\ell$.

\vspace{-0.25cm}
\subsubsection{Composite scenario}
\vspace{-0.15cm}
In the composite-fermion scenario, quarks and leptons are assumed to have an internal substructure that should manifest itself at some sufficiently high energy scale, the compositeness scale $\Lambda$.
Ordinary fermions are considered as bound states of some not-yet observed fundamental constituents generically referred to as $preons$.
Two model-independent properties are experimentally relevant: the existence of a contact interaction, in addition to the gauge
interaction, which represents an effective approach for describing the effects of the unknown internal dynamics of compositeness, and the existence of excited states of quarks and leptons
with masses lower than or equal to $\Lambda$. A particular case of such excited states could be a heavy composite Majorana neutrino (HCMN).
The production cross-section depends on $m_{N_{\ell}}$ and $\Lambda$.

\vspace{-0.25cm}
\subsection{Microscopic black holes}
\vspace{-0.15cm}
Models of TeV-scale gravity, like ADD \cite{ADD} or RS \cite{RS}, postulate that the fundamental scale of gravity, $M_{D}$, in a higher-dimensional
space-time is much lower than that measured in our four-dimensional space-time.
Interesting signatures are expected in these models in the form of
non-perturbative gravitational states such as microscopic black holes, which are assumed to be produced over a continuous range of mass values above a threshold, $M_{th}$.
They are investigated below in the final state with one lepton and either two additional leptons or two jets.

In the following, we summrize searches from the ATLAS and CMS collaborations.
A complete description of the objects used in the analyses described below can be found in the corresponding references.

\vspace{-0.5cm}
\section{Search for dielectron plus dijet or dimuon plus dijet signatures}
\vspace{-0.15cm}
Dielectron plus dijet and dimuon plus dijet signatures are investigated in the search for first- and second-generation leptoquarks and LR models.
Results are shown for measurements with data collected in $pp$ collisions at $\sqrt{s}$ = 13 TeV. 
The ATLAS collaboration uses 3.2 fb$^{-1}$ and looks for a leptoquark pair \cite{LQ_ATLAS}, while 
the CMS collaboration uses 2.6 fb$^{-1}$ and looks for a leptoquark pair \cite{LQ1_CMS,LQ2_CMS} and $W_R$ and $N_{\ell}$ from the LR model \cite{LR_CMS}. In these analyses no dilepton charge requirement is demanded.
Same-sign dilepton plus dijet signatures are instead investigated 
by the ATLAS \cite{lljj8TeV_ATLAS} and CMS \cite{lljj8TeV_CMS1}, \cite{lljj8TeV_CMS2} collaborations, using about 20 fb$^{-1}$ of data collected in $pp$ collisions at $\sqrt{s}$ = 8 TeV.
Results are interpreted in terms of the LR model parameters and in the context of a Type-I seesaw mechanism and are summarized in Sec. \ref{Summary}.

\vspace{-0.25cm}
\subsection{Searches for first- and second-generation leptoquarks in ATLAS}
\begin{figure}[h!]
\subfloat[]{\label{LQ1_Mmlq_ATLAS}\includegraphics[width = 3.4 cm, height = 3.25 cm]{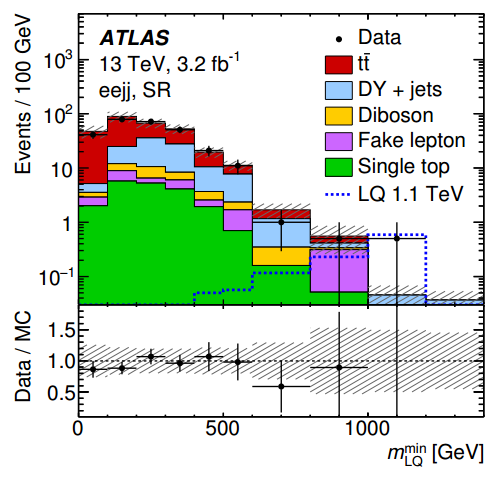}} 
\hspace{0.75cm}
\subfloat[]{\label{LQ2_Mmlq_ATLAS}\includegraphics[width = 3.4 cm, height = 3.25 cm]{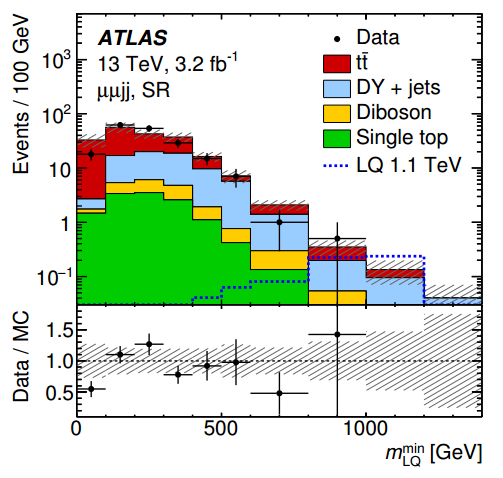}}\
\hspace{0.75cm}
\subfloat[]{\label{LQ1_Lim_ATLAS}\includegraphics[width = 3.4 cm, height = 3.25 cm]{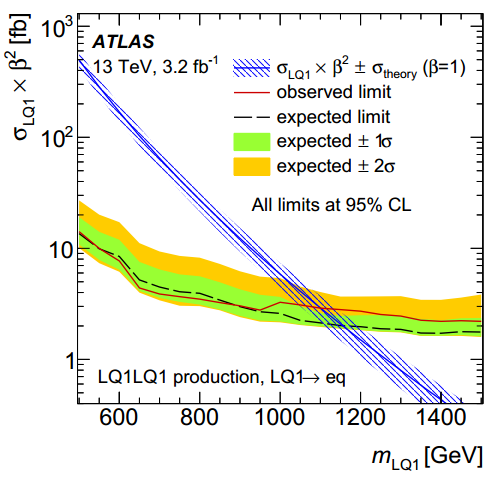}} 
\hspace{0.75cm}
\subfloat[]{\label{LQ2_Lim_ATLAS}\includegraphics[width = 3.4 cm, height = 3.25 cm]{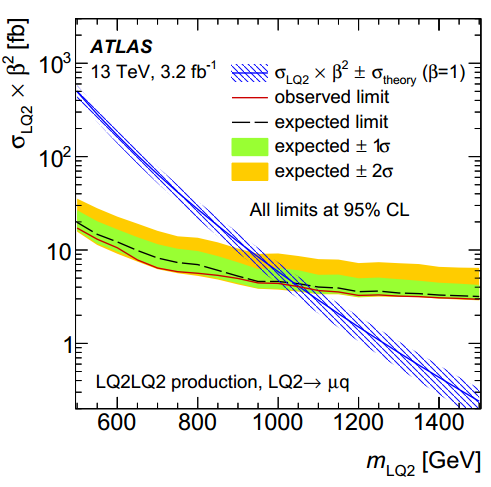}}\
\vspace{-0.35cm}
\caption{ATLAS results of searches for first- and second-generation leptoquarks \cite{LQ_ATLAS}: $m^{\mathrm{min}}_{LQ}$ distribution in the eejj (a) and $\mu\mu$jj (b) channels,
95\% CL exclusion limits on $M_{LQ}$ from the eejj (c) and $\mu\mu$jj (d) channels.}
\label{LQ_ATLAS}
\end{figure}

\vspace{-0.15cm}
The signal selection is defined requiring exactly two leptons with electron (muon) transverse momentum $p_\mathrm{T}$ $>$ 30 (40) GeV, $|\eta| <$ 2.5 and at least two jets with $p_\mathrm{T}$ $>$ 40 GeV, $|\eta| <$ 2.5,
$m_{\ell\ell} >$ 130 GeV, and $S_\mathrm{T}$, the scalar sum of the transverse momentum of the two leptons and of the two leading jets, greater than 600 GeV.
$m^{\mathrm{min}}_{LQ}$, the minimum invariant mass of the two lepton-jet pairs that minimize their mass difference, is used for the signal extraction.
Major backgrounds arise from DY+jets and $t\bar{t}$ processes.
Both are estimated by normalizing the Monte Carlo (MC) simulation by scale factors measured around the Z boson mass (70 $< m_{\ell\ell} <$ 110 GeV)
and in a control region with at least two jets, and exactly one muon and one electron.
No excess of events compared to the SM estimations is observed, as it can be seen from Figs.~\ref{LQ1_Mmlq_ATLAS},~\ref{LQ2_Mmlq_ATLAS}. 
95\% confidence level (CL) upper bounds on the scalar leptoquark pair production cross section times branching ratio of decay in the eejj and $\mu\mu$jj channels are determined,
taking into account all the systematics described in Ref. \cite{LQ_ATLAS}. 
Leptoquarks are excluded for masses $M_{LQ} <$ 1100 (1050) GeV in the ee($\mu\mu$)jj channels, assuming $\beta$ = BR(LQ $\rightarrow \ell$q ) = 1, as reported in Figs.~\ref{LQ1_Lim_ATLAS},~\ref{LQ2_Lim_ATLAS}.
\vspace{-0.35cm}

\vspace{-0.75cm}
\subsection{Searches for first- and second-generation leptoquarks in CMS}
\begin{figure}[h]
\subfloat[]{\label{LQ1_Mmlq_CMS}\includegraphics[width = 3.5 cm, height = 3. cm]{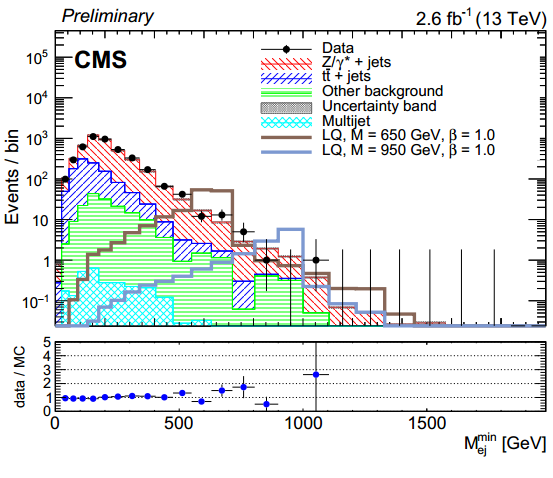}}
\hspace{0.5cm}
\subfloat[]{\label{LQ2_Mmlq_CMS}\includegraphics[width = 3.5 cm, height = 3. cm]{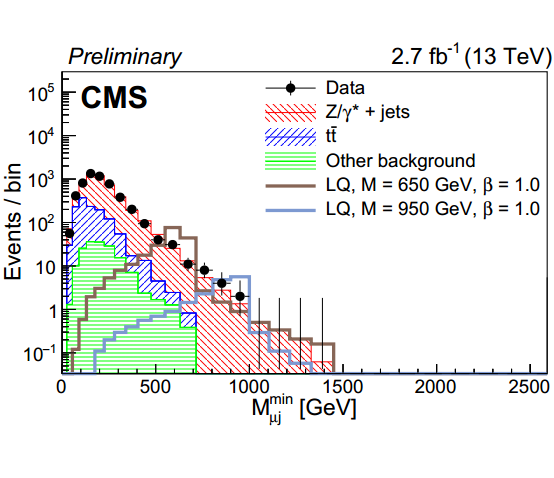}}\
\hspace{0.5cm}
\subfloat[]{\label{LQ1_Lim_CMS}\includegraphics[width = 3.5 cm, height = 3. cm]{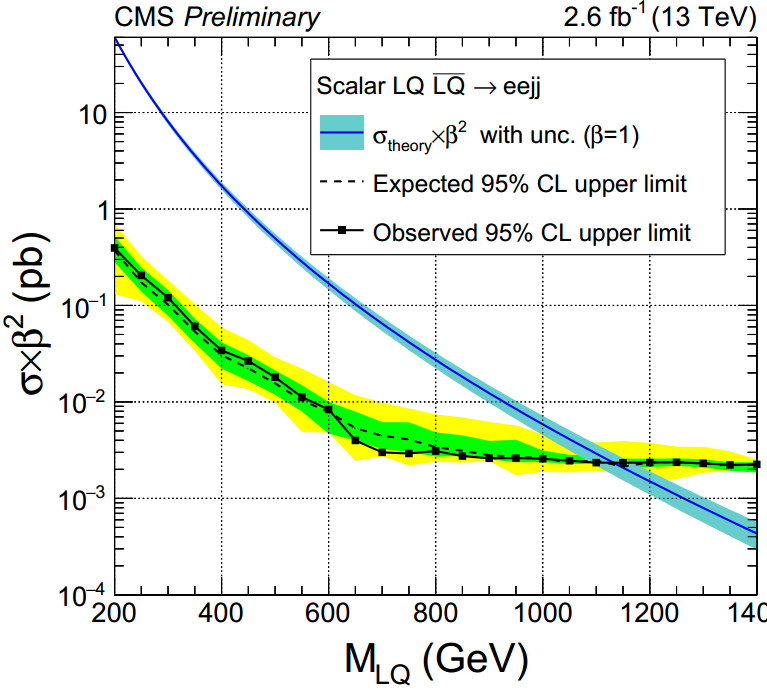}}
\hspace{0.5cm}
\subfloat[]{\label{LQ2_Lim_CMS}\includegraphics[width = 3.5 cm, height = 3. cm]{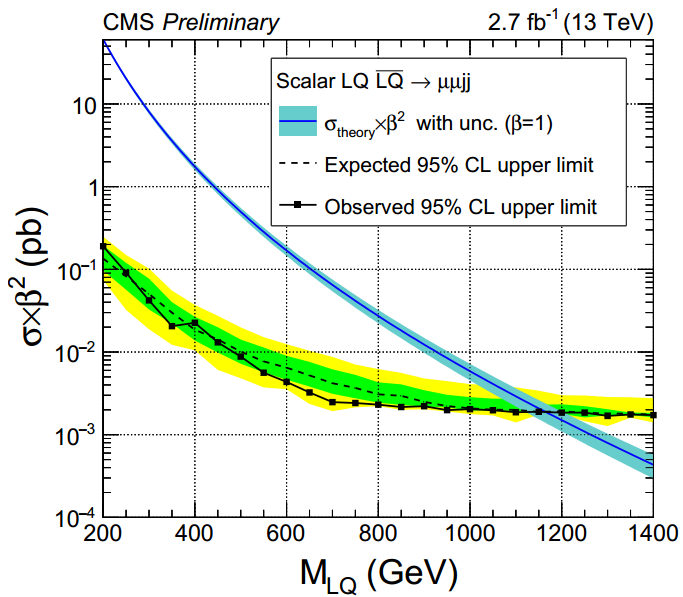}} 
\vspace{-0.35cm}
\caption{
CMS results of searches for first- and second-generation leptoquarks \cite{LQ1_CMS,LQ2_CMS}: $M^{\mathrm{min}}_{ej,\mu j}$ distribution in the eejj (a) and $\mu\mu$jj (b) channels,
95\% CL exclusion limits on $M_{LQ}$ from the eejj (c) and $\mu\mu$jj (d) channels.}
\label{some example}
\end{figure}

\vspace{-0.15cm}
Events are selected requiring precisely two leptons with electron (muon) $p_\mathrm{T}$ $>$ 50 GeV, $|\eta| <$ 2.5 (2.1) and at least two jets with $p_\mathrm{T}$ $>$ 50 GeV, $|\eta| <$ 2.4.
The signal region is complemented by the selection on three variables, $m_{\ell\ell}$, $S_\mathrm{T}$, and $m^{\mathrm{min}}_{LQ}$, which is optimized for each LQ mass hypothesis to achieve the highest sensitivity.
The main backgrounds, DY+jets and $t\bar{t}$, are estimated by adopting the same techniques used by ATLAS,
except for the $t\bar{t}$ contamination in the second-generation LQ analysis, which 
is evaluated from data events selected with an electron and a muon rescaled
to account for differences in the selection of the e$\mu$ and and $\mu\mu$ final states.
A good agreement is observed between the data and the SM estimations, as shown in Figs.~\ref{LQ1_Mmlq_CMS},~\ref{LQ2_Mmlq_CMS}. 
95\% CL upper limits on the scalar leptoquark pair production cross section times branching ratio of decay in the eejj and $\mu\mu$jj channels is calculated,
accounting for all the systematics described in Refs. \cite{LQ1_CMS,LQ2_CMS}. 
Leptoquarks are excluded for masses $M_{LQ} <$ 1130 (1165) GeV in the ee($\mu\mu$)jj channels, assuming $\beta$ = BR(LQ $\rightarrow \ell$q ) = 1, as reported in Figs.~\ref{LQ1_Lim_CMS},~\ref{LQ2_Lim_CMS}.
\vspace{-0.25cm}
\subsection{Searches for $W_R$ and right-handed neutrinos in CMS}
\begin{figure}[h]
\subfloat[]{\label{WRee_Mass_CMS}\includegraphics[width = 3.069 cm, height = 3. cm]{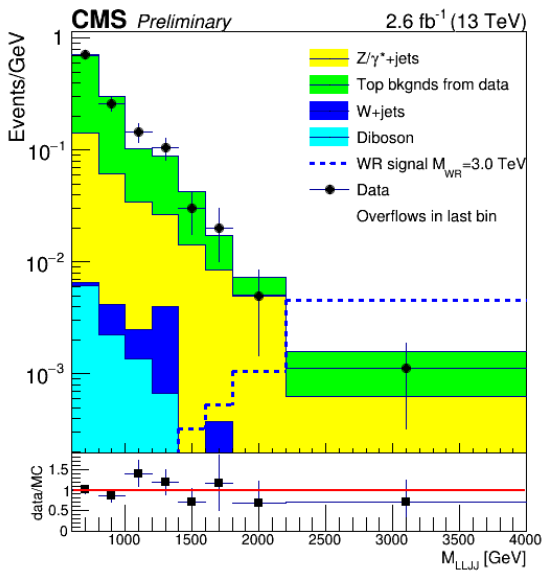}}
\hspace{1.1cm}
\subfloat[]{\label{WRmm_Mass_CMS}\includegraphics[width = 3.069 cm, height = 3. cm]{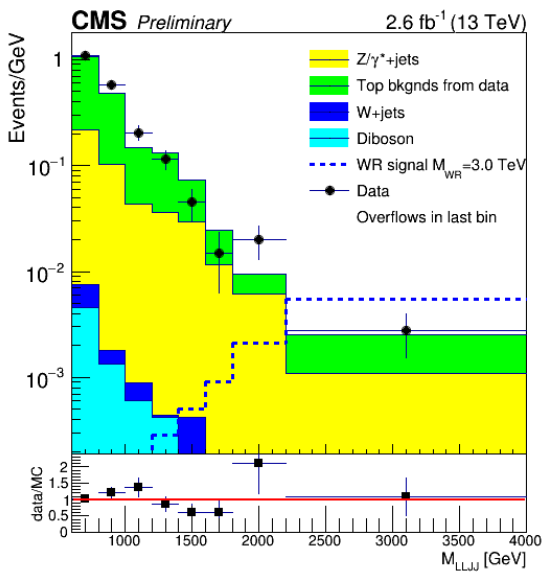}}\
\hspace{1.1ccm}
\subfloat[]{\label{WRee_Lim2_CMS}\includegraphics[width = 3.069 cm, height = 3. cm]{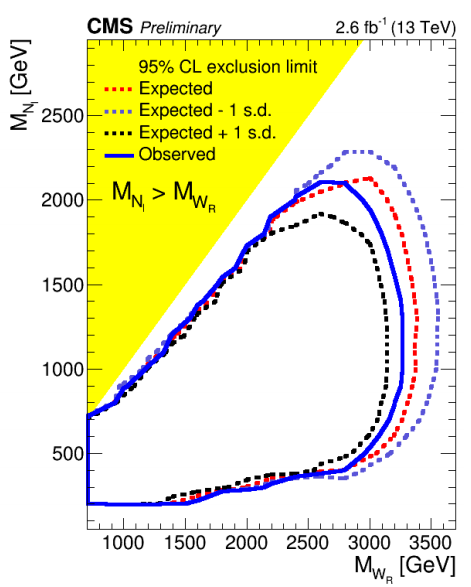}}
\hspace{1.1cm}
\subfloat[]{\label{WRmm_Lim2_CMS}\includegraphics[width = 3.069 cm, height = 3. cm]{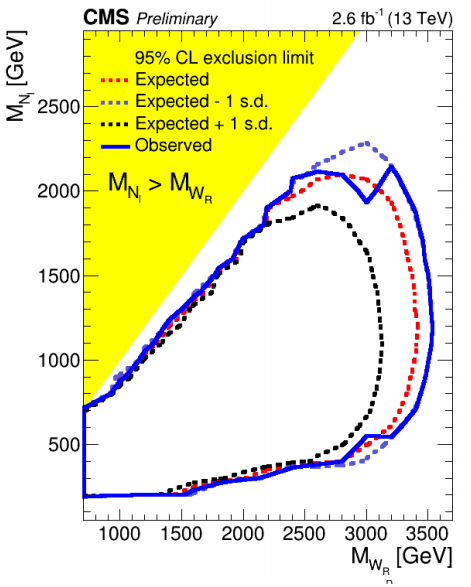}} 
\vspace{-0.35cm}
\caption{
CMS results of searches for $W_R$ and right-handed neutrinos \cite{LR_CMS}: $M_{\ell\ell jj}$ distribution in the eejj (a) and $\mu\mu$jj (b) channels,
95\% CL exclusion limits on $M_{LQ}$ from the eejj (c) and $\mu\mu$jj (d) channels.
}
\label{some example}
\end{figure}

\vspace{-0.15cm}
Two leptons with electron (muon) transverse momentum $p_\mathrm{T}$ $>$ 60 (53) GeV, $|\eta| <$ 2.4 and at least two jets with $p_\mathrm{T}$ $>$ 40 GeV, $|\eta| <$ 2.4 are sought,
together with $m_{\ell\ell} >$ 200 GeV and $M_{\ell\ell jj} >$ 600 GeV.
The main backgrounds, DY+jets and $t\bar{t}$, are estimated using the same techniques implemented in searches for second-generation leptoquarks in CMS.
The data and the SM expectations in the signal region are in good agreement, as it is illustrated in Figs.~\ref{WRee_Mass_CMS},~\ref{WRmm_Mass_CMS}.
95\% CL exclusion limits in the ($M_{W_R}$, $M_{N_{\ell}}$) plane for the process $pp \rightarrow W_R \rightarrow \ell N_{\ell} \rightarrow \ell \ell$jj ($\ell$ = e,$\mu$) are evaluated,
considering all the systematics described in Ref. \cite{LR_CMS}, and shown in Figs.~\ref{WRee_Lim2_CMS},~\ref{WRmm_Lim2_CMS}.
\vspace{-0.35cm}

\vspace{-0.75cm}

\vspace{-1.cm}
\section{Search for ditau plus dijet signature}
\vspace{-0.15cm}
Ditau plus dijet signature is investigated in the search for third-generation leptoquarks and $W_R$, $N$ from the LR model,
using data from $pp$ collisions recorded by the CMS experiment at $\sqrt{s}$ = 13 TeV.
2.1 fb$^{-1}$ \cite{LQ3015_CMS} and 12.9 fb$^{-1}$ \cite{LQ3016_CMS} are used in analyses that consider full hadronic and semileptonics ditau decay channels.

\begin{figure}[h!]
\subfloat[]{\label{LQ3_CMS_ST_1.png}\includegraphics[width = 3. cm, height = 2.75 cm]{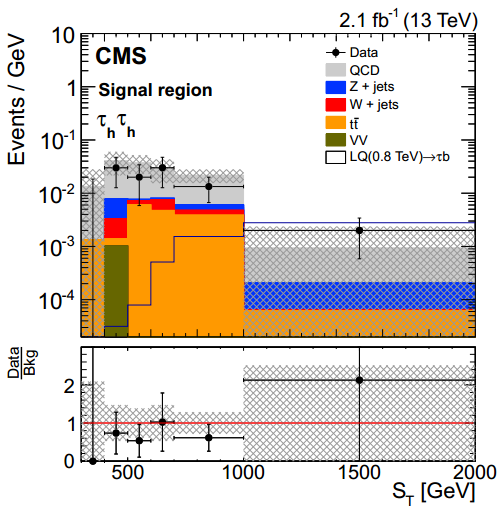}}
\hspace{0.15cm}
\subfloat[]{\label{LQ3_CMS_ST_2.png}\includegraphics[width = 3. cm, height = 2.75 cm]{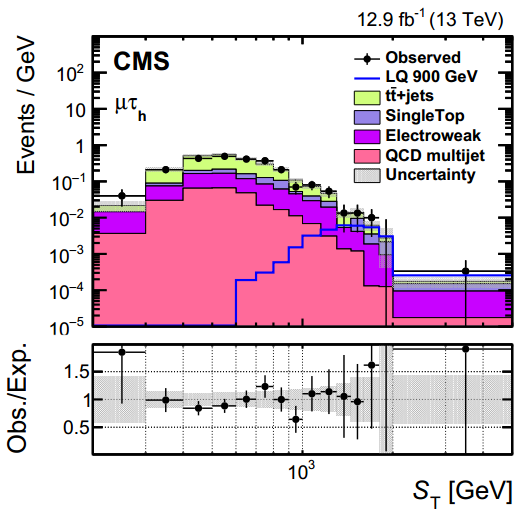}}\
\hspace{0.15cm}
\subfloat[]{\label{LQ3_CMS_ST_3.png}\includegraphics[width = 3. cm, height = 2.75 cm]{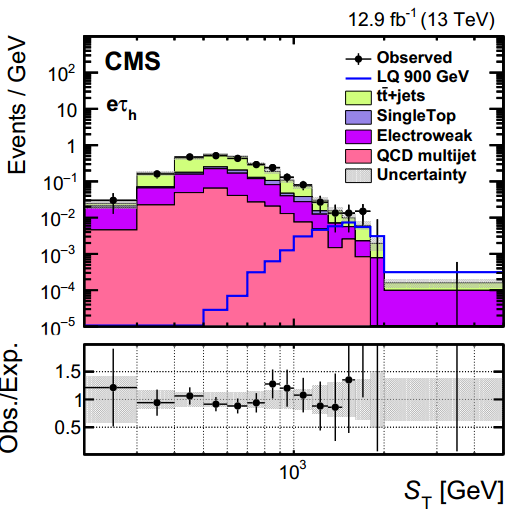}}
\hspace{0.15cm}
\subfloat[]{\label{WRtauhtauh_Lim_CMS.png}\includegraphics[width = 3. cm, height = 2.75 cm]{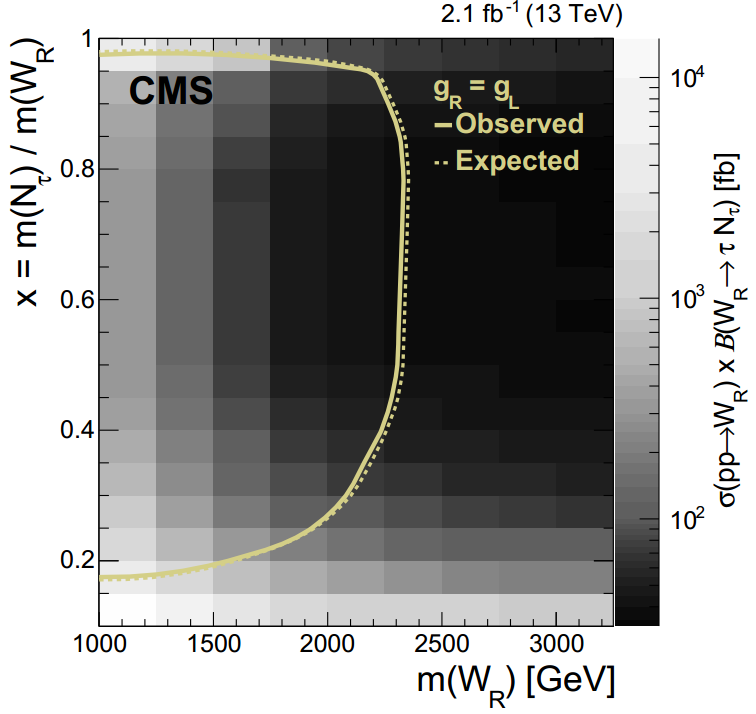}} 
\hspace{0.15cm}
\subfloat[]{\label{WRtauhl_Lim_CMS.png}\includegraphics[width = 3.3 cm, height = 2.75 cm]{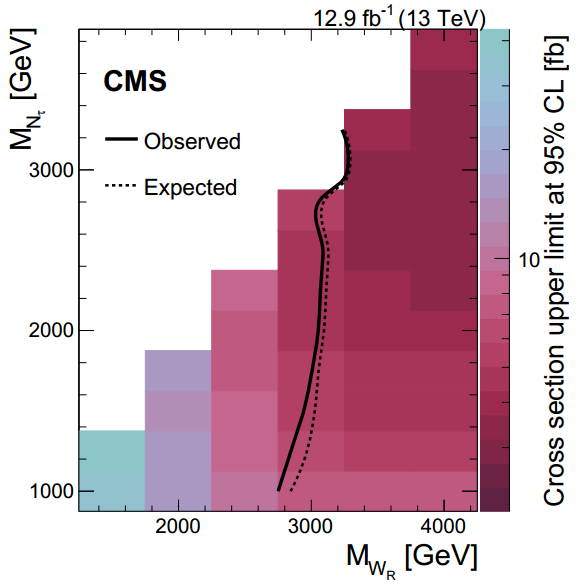}} 
\vspace{-0.35cm}
\caption{
CMS results of searches for $W_R$ and hight-handed neutrinos \cite{LQ3015_CMS}, \cite{LQ3016_CMS}: $S_\mathrm{T}$ distribution in the $\tau_h\tau_h$jj (a), $\mu \tau_h$jj (b), e$\tau_h$jj (c) channels,
95\% CL 2D exclusion limits from the $\tau_h \tau_h$jj (d) and the combined e,$\mu$ plus  $\tau_h$jj (e) channels.}
\label{some example}
\end{figure}

Signal events are considered if they have either two hadronic taus ($\tau_h$) with $p_\mathrm{T}$ $>$ 70 GeV, $|\eta|$ $<$ 2.1, or a $\tau_h$ plus an electron or a muon with $p_\mathrm{T}$ $>$ 50 GeV, $|\eta|$ $<$ 2.1,
paired to 2 jets with $p_\mathrm{T}$ $>$ 50 GeV, $|\eta|$ $<$ 2.4.
Missing transverse energy, $E^{miss}_{T}$, greater than 50 GeV and the invariant mass of the ditau visible decay products greater than 100 GeV are further sought. 
The dominant reducible backgrounds are given by QCD multijet events, in the full hadronic channel, and W+jets, in the semileptonic channels, are estimated with fully data-driven techniques. 
The other backgrounds are estimated using MC simulation, after validation of the good modeling for the relevant processes in dedicated control regions.
Data are found to be in agreement with the SM expectation, as it is shown in Figs.~\ref{LQ3_CMS_ST_1.png},~\ref{LQ3_CMS_ST_2.png},~\ref{LQ3_CMS_ST_3.png}.
95\% CL exclusion limits in the ($M_{W_R}$, $M_{N_{\ell}}$) plane for the process $pp \rightarrow W_R \rightarrow \ell N_{\ell} \rightarrow \ell \ell$jj ($\ell$ = $\tau$) are evaluated,
including all the systematics described in Refs.~\cite{LQ3015_CMS},~\cite{LQ3016_CMS} and shown in Figs.~\ref{WRtauhtauh_Lim_CMS.png},~\ref{WRtauhl_Lim_CMS.png}.

\vspace{-0.4cm}

\vspace{0.5cm}
\section{Search for dilepton plus one large-radius jet signature}
\vspace{-0.15cm}
Dilepton (electron or muon) plus diquark signature is investigated in search for heavy composite Majorana neutrinos \cite{HCMN_CMS},
using 2.3 fb$^{-1}$ of data picked up by the CMS experiment from $pp$ collisions at $\sqrt{s}$ = 13 TeV.

\begin{figure}[h!]
\subfloat[]{\label{M_leplepBjet_SRe}\includegraphics[width = 3.93 cm, height = 3. cm]{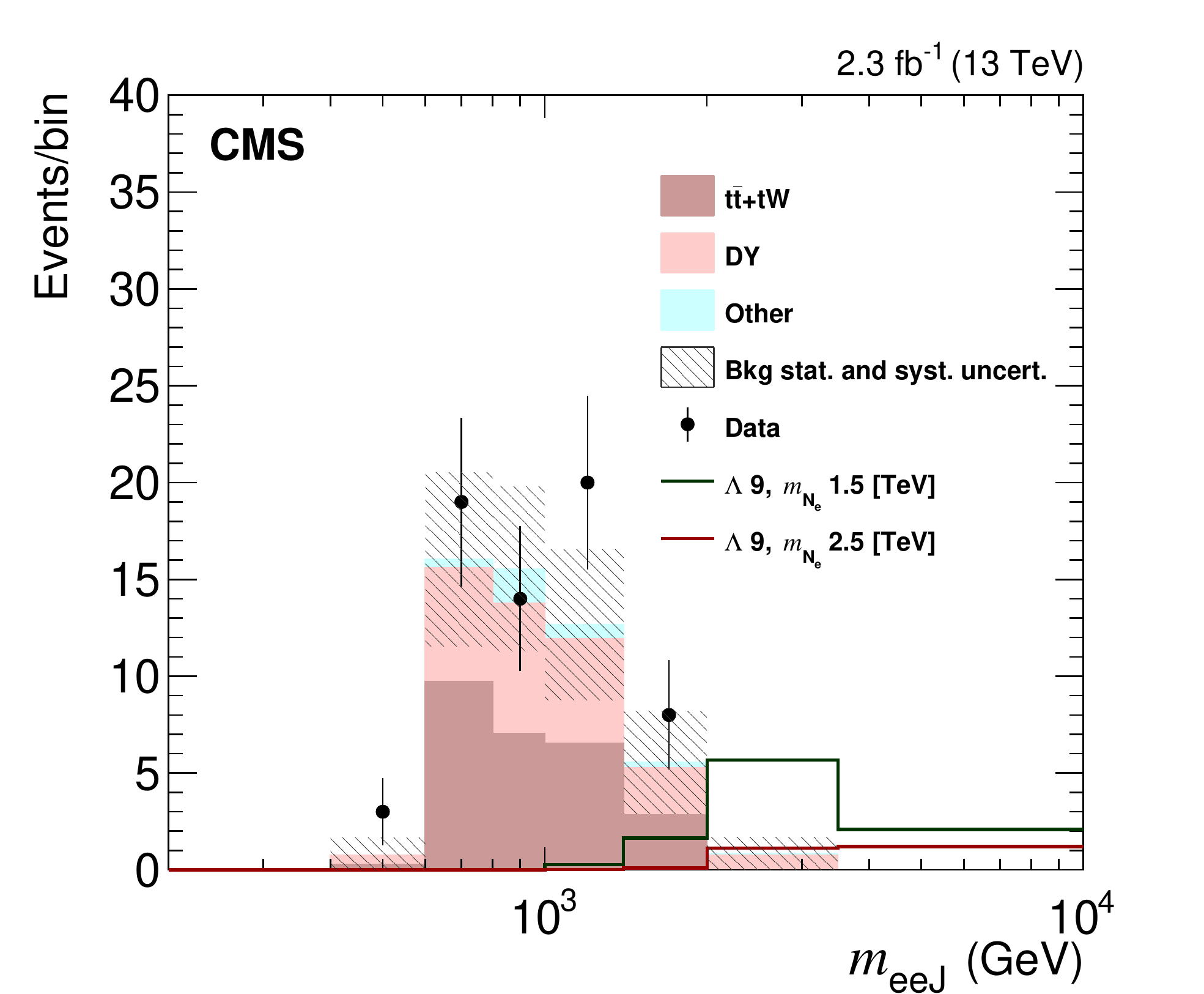}}
\hspace{0.15cm}
\subfloat[]{\label{M_leplepBjet_SRmu}\includegraphics[width = 3.93 cm, height = 3. cm]{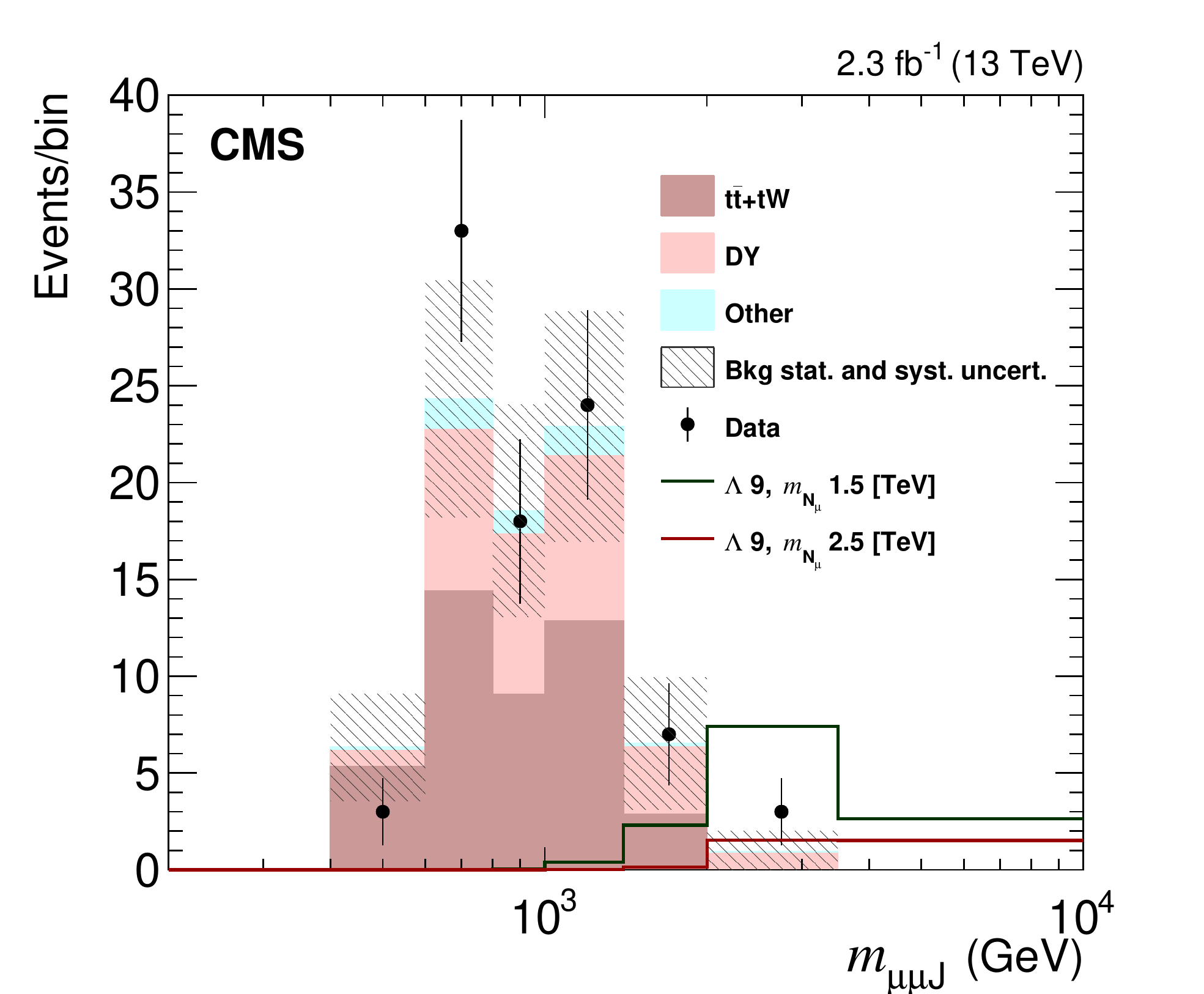}}\
\hspace{0.15cm}
\subfloat[]{\label{eejj_2D_limit}\includegraphics[width = 3.93 cm, height = 3. cm]{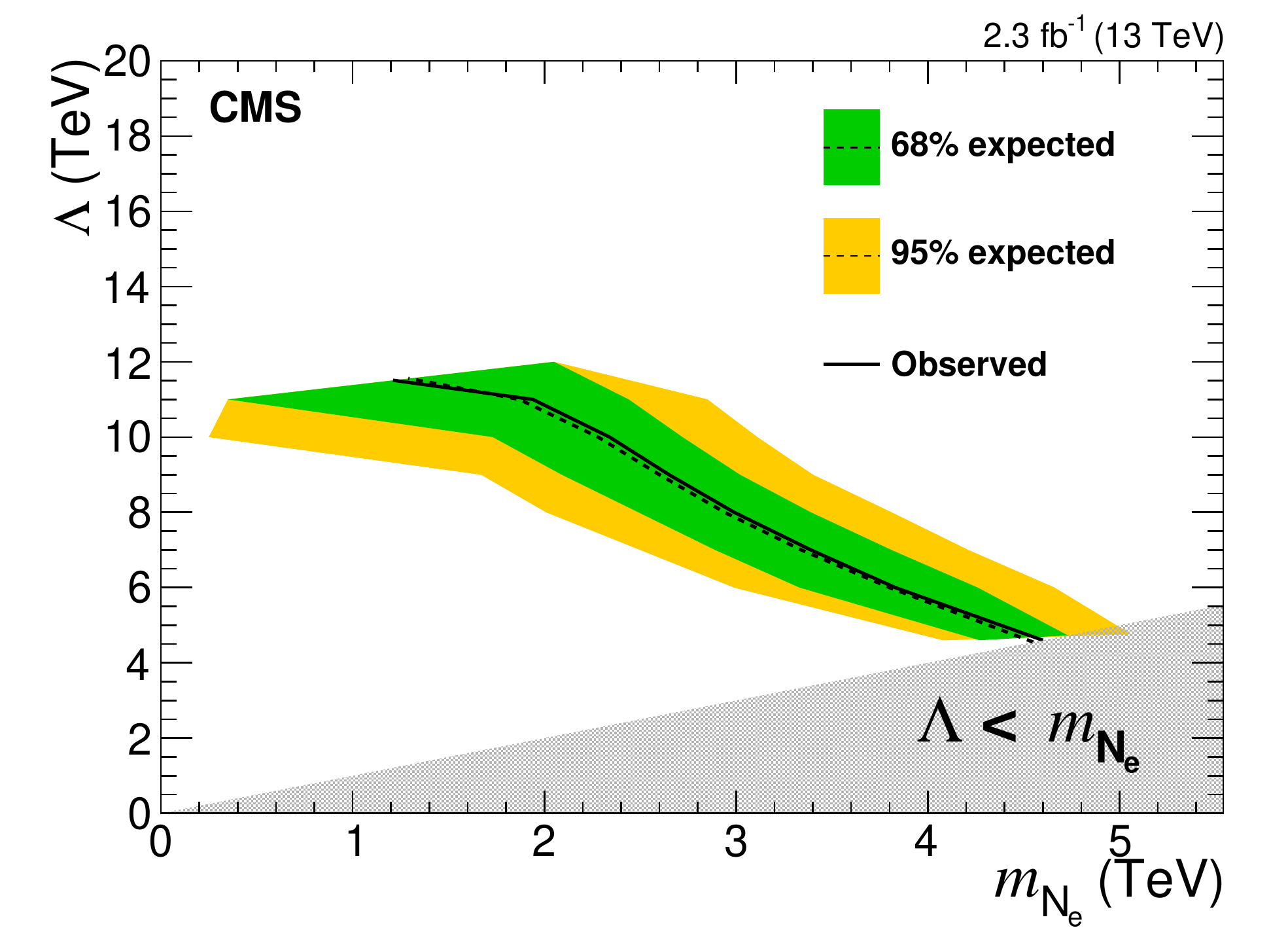}}
\hspace{0.15cm}
\subfloat[]{\label{mumujj_2D_limit}\includegraphics[width = 3.93 cm, height = 3. cm]{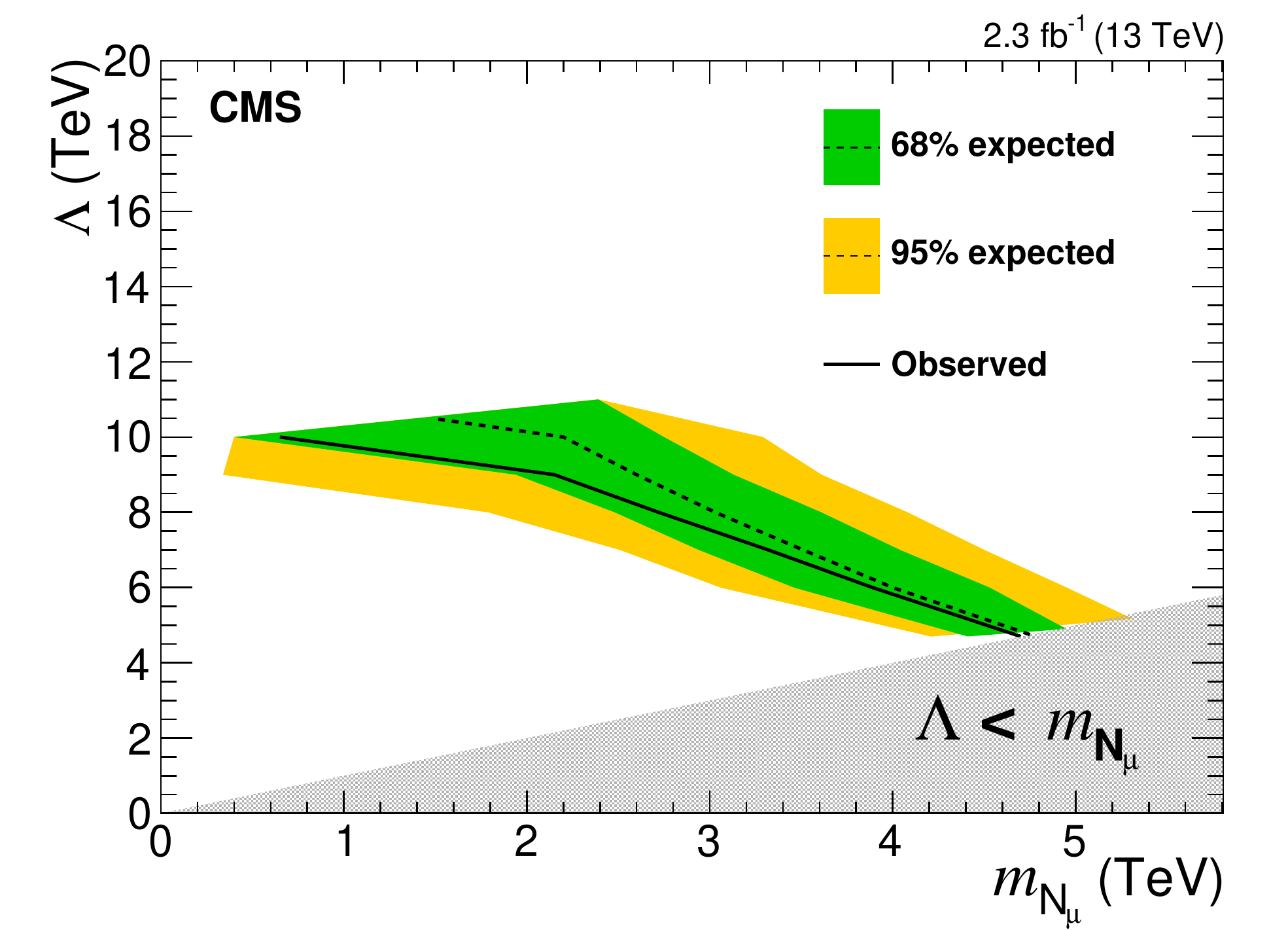}} 
\vspace{-0.35cm}
\caption{CMS results of searches for heavy composite Majorana neutrinos \cite{HCMN_CMS}: $M_{eeJ}$ (a) and $M_{\mu\mu J}$ (b) distributions,
95\% CL 2D exclusion limits from the eejj (c) and $\mu\mu$jj (d) channels.
}
\label{some example}
\end{figure}

Events entering the signal region are characterized by the presence of two electrons, with $p_\mathrm{T}$ $>$ 100 and 35 GeV, 
or two muons, with $p_\mathrm{T}$ $>$ 50 and 30 GeV, and one large-radius jet (anti-$k_{\rm t}$ \cite{Cacciari:2008gp} with radius equal to 0.8), with $p_\mathrm{T}$ $>$ 190 GeV, being $|\eta| <$ 2.4 for all the objects.
The large-radius jet is chosen to keep high signal acceptance whichever is the interaction, gauge or contact, that rules the process under investigation $pp \rightarrow \ell N_{\ell} \rightarrow \ell\ell q \bar{q}'$.
The main contamination comes from DY+jets and $t\bar{t}$ processes, which are estimated using techniques similar to those used in searches for second-generation leptoquarks in CMS.
The data and the SM expectations in the signal region are in good agreement, as is illustrated in Figs.~\ref{M_leplepBjet_SRe},~\ref{M_leplepBjet_SRmu}.
95\% CL exclusion limits in the ($\Lambda$, $M_{N_{\ell}}$) plane for the process $pp \rightarrow \ell N_{\ell} \rightarrow \ell \ell$jj ($\ell$ = e,$\mu$) are evaluated,
accounting for all the systematics described in Ref. \cite{HCMN_CMS}, and shown in Figs.~\ref{eejj_2D_limit},~\ref{mumujj_2D_limit}.

\vspace{-0.35cm}

\vspace{-0.75cm}
\section{Search for one lepton and two jets or two more leptons signature}
\vspace{-0.15cm}
The one lepton and two jets or two more leptons signature is studied in searches for models of microscopic black holes \cite{BH_ATLAS},
using 3.2 fb$^{-1}$ of $pp$ collision data taken by the ATLAS detector at $\sqrt{s}$ = 13 TeV.

\begin{figure}[h!]
\subfloat[]{\label{GravityModels_1}\includegraphics[width = 3.37 cm, height = 3. cm]{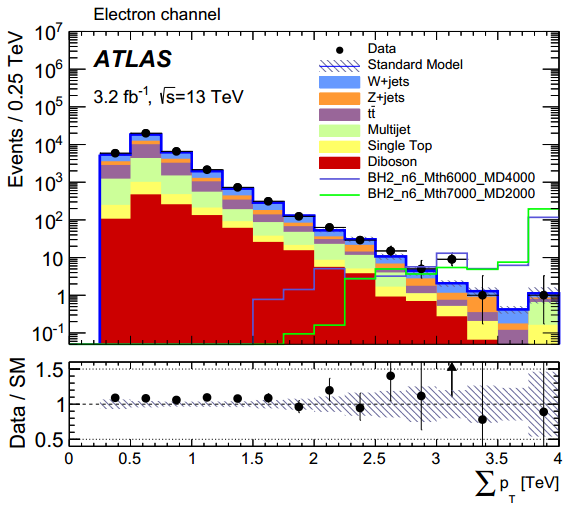}}
\hspace{2.75cm}
\subfloat[]{\label{GravityModels_2}\includegraphics[width = 3.37 cm, height = 3. cm]{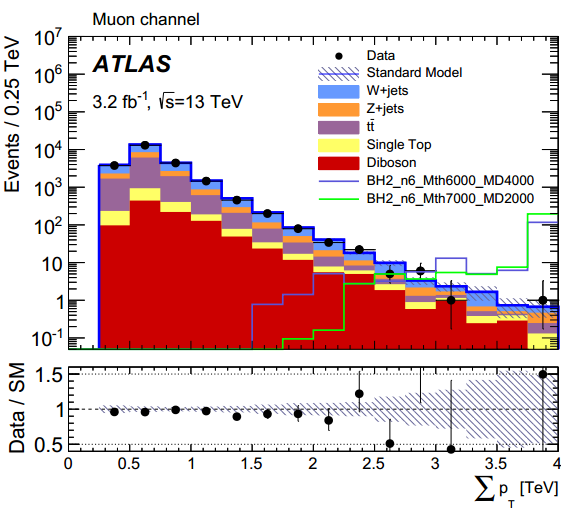}}\
\hspace{2.75cm}
\subfloat[]{\label{GravityModels_3}\includegraphics[width = 3.37 cm, height = 3. cm]{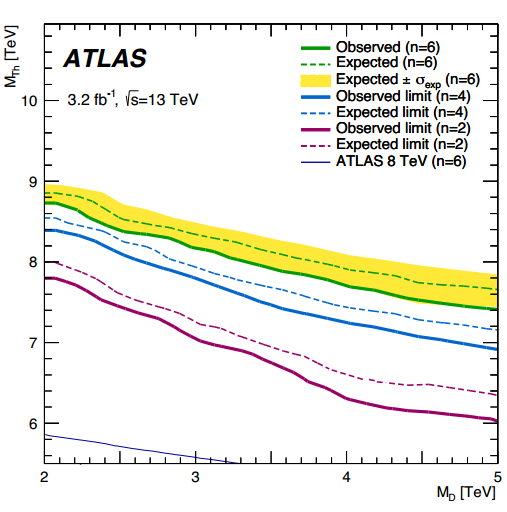}}
\vspace{-0.35cm}
\caption{ATLAS results of searches for microscopic black holes \cite{BH_ATLAS}: $\sum$ p$_{\scriptsize\textrm{T}}$ distributions in the electron (a) and muon (b) channels,
95\% CL 2D exclusion limits (c).
}
\label{some example}
\end{figure}

The search is performed selecting leptons and jets with $p_\mathrm{T}$ $>$ 100 GeV and the sum of the transverse momenta of the final state objects, $\sum$ p$_\mathrm{T}$, greater than 2 TeV.
The main backgrounds, Z,W+jets and $t\bar{t}$, are estimated from the MC simulation normalized to data, using dedicated control regions.
The data and the SM expectations in the signal region are in good agreement, as is illustrated in Figs.~\ref{GravityModels_1},~\ref{GravityModels_2}.
95\% CL exclusion limits in the (M$_{D}$, M$_{th}$) plane are evaluated,
considering all the systematics described in Ref. \cite{BH_ATLAS}, and shown in Fig.~\ref{GravityModels_3}.

\vspace{-0.35cm}

\vspace{-0.35cm}
\section{Summary}
\label{Summary}
\vspace{-0.15cm}
In this proceeding we have summarized the most recent results at both ATLAS and CMS collaborations of searches in lepton plus jet final states 
motivated by models dealing with leptoquarks, heavy Majorana neutrinos, and microscopic black holes.
In all measurements, the data are in good agreement with the SM prediction and limits are set on the parameters of the different models.
95\% CL exclusion limits on leptoquark mass are summarized in Table 1, while the results of heavy Majorana neutrinos searches are in Table 2. 
The production of microscopic black holes is also investigated and excluded in models with two to six extra space dimensions in the (M$_D$, M$_{th}$) plane, as reported in Fig. 6c.

\begin{table}[h!]
\begin{center}
{\scriptsize 
\begin{tabular}{|c|c|c|c|}
\hline
               & LQ1                       & LQ2                            & LQ3 \\
\hline
 ATLAS         & $<$1100 GeV ($ee+jj$)     & $<$1050 GeV ($\mu\mu+jj$)      &  $<$640 GeV ($t\bar{t}+E\mathrm{^{miss}_T}$, 8 TeV)     \\
\hline
 CMS           & $<$1130 GeV ($ee+jj$)     & $<$1165 GeV ($\mu\mu+jj$)      &  $<$740 GeV ($\tau_h \tau_h+bb$)     \\
               &                           &                                &  $<$850 GeV ($\tau_h \ell +bb$)      \\
\hline
\end{tabular}
}
\end{center}
\label{LQsummary}
\vspace{-0.35cm}
\caption{Summary of ATLAS and CMS exclusion limits for the leptoquark mass with respect to the three leptoquark generations (Refs.~\cite{LQ_ATLAS},~\cite{LQ1_CMS},~\cite{LQ2_CMS},~\cite{LQ3015_CMS},~\cite{LQ3016_CMS}).}
\end{table}  

\vskip-5pt
{\tiny  
\begin{table}[h!]
\begin{center}
{\scriptsize 
\begin{tabular}{|c|c|c|c|}
\hline
               & Left$-$right                  & Type I seesaw (8 TeV)        & Composite \\
\hline
 ATLAS         & 50 GeV to 2000 GeV          & $|V_{eN}|^2<$ 0.029            & n/a\\
               & $m_{W_R}>$ 400 GeV          & $|V_{\mu N}|^2<$ 0.0028        & \\
               &                             & for $m_N$ = 110  GeV           & \\ 
\hline
 CMS           & 200 GeV to 2150 GeV         & $|V_{eN}|^2<$ 0.00015-0.71                                                  &  4.35 $ee qq$  TeV \\
               & $m_{W_R}>$ 600 GeV          & $|V_{\mu N}|^2<$ 2.1$\times$10$^{-5}$-0.583                                 &  4.50 $\mu\mu qq$ TeV\\
               &                             & $|V_{eN}V_{\mu N}^*/(|V_{eN}|^2+|V_{\mu N}|^2)|<$ 6.6$\times$10$^{-5}$-0.47 &  for $\Lambda$ = 5 TeV\\
               &                             & for $m_N$ in [40-500] GeV                                                   & \\ 
\hline
\end{tabular}
\label{HMNsummary}
}
\end{center}
\vspace{-0.35cm}
\caption{Summary of ATLAS and CMS exclusion limits for the parameters of the models with heavy Majorana neutrinos discussed in this proceeding (Refs.~\cite{LR_CMS},~\cite{lljj8TeV_ATLAS},~\cite{lljj8TeV_CMS1},~\cite{lljj8TeV_CMS2},~\cite{HCMN_CMS}).}
\end{table}  

\vspace{-0.25cm}
\bibliographystyle{plain}
\bibliography{2017_LHCP_Proceeding_FR}

\end{document}